%% file: main.tex
\documentclass[10pt,conference]{IEEEtran}
\IEEEoverridecommandlockouts
\pdfoutput=1
\usepackage{graphicx} 
\usepackage[T1]{fontenc}
\usepackage{cite}
\usepackage{amsmath}
\usepackage{subcaption}
\usepackage{ulem}
\usepackage[usenames,dvipsnames]{xcolor}
\usepackage{textcomp}
\usepackage{dcolumn} 
\usepackage{bm} 
\usepackage{braket}
\usepackage{hyperref} 
\hypersetup{
    colorlinks = true,
    citecolor = magenta,
    linkcolor = purple
}

\title{Determining Molecular Ground State with Quantum Imaginary Time Evolution using Broken-Symmetry Wave Function \\
\thanks{This work was supported by Quantum-LEAP Flagship Program (Grant No. JPMXS0120319794) from Ministry of Education, Culture, Sports, Science and Technology (MEXT), Japan and Japan Science and Technology Agency (JST) [Moonshot R\&D] (JPMJMS226C) and (JPMJMS2061). K. S. acknowledges support from Center of Innovations for Sustainable Quantum AI (JPMJPF2221) from Japan Science and Technology Agency (JST), Japan, and Grants-in-Aid for Scientific Research C (21K03407) and for Transformative Research Area B (23H03819) from Japan Society for the Promotion of Science (JSPS), Japan.}
}

\author{
\IEEEauthorblockN{
Pawan Sharma Poudel\IEEEauthorrefmark{1}\IEEEauthorrefmark{4}\thanks{\IEEEauthorrefmark{4}These authors contributed equally to this work.},
Kenji Sugisaki\IEEEauthorrefmark{3}\IEEEauthorrefmark{4},
Michal Hajdu\v{s}ek\IEEEauthorrefmark{1},
and Rodney Van Meter\IEEEauthorrefmark{2}}\\
\IEEEauthorblockA{\IEEEauthorrefmark{1}\textit{Graduate School of Media and Governance, Keio University, 5322 Endo, Fujisawa, Kanagawa 252-0882, Japan}}
\IEEEauthorblockA{\IEEEauthorrefmark{2}\textit{Faculty of Environment and Information Studies, Keio University, 5322 Endo, Fujisawa, Kanagawa 252-0882, Japan}
\IEEEauthorblockA{\IEEEauthorrefmark{3}\textit{Quantum Computing Center, Keio University, 3-14-1 Hiyoshi, Kohoku-ku, Yokohama, Kanagawa 223-8522, Japan}\\
 \{pawan, michal, rdv\}@sfc.wide.ad.jp, ksugisaki@keio.jp}
}}

\date{February 2025}

\input{custom-commands}

\begin{document}
\bstctlcite{IEEEexample:BSTcontrol}

\maketitle
\thispagestyle{plain}
\pagestyle{plain} 

\begin{abstract}
The Hartree--Fock (HF) wave function, commonly used for approximating molecular ground states, becomes nonideal in open shell systems due to the inherent multi-configurational nature of  the wave function, limiting accuracy in Quantum Imaginary Time Evolution (QITE). We propose replacing the HF wave function with a spin- and spatial-symmetry broken wave function, enhancing convergence by adding a spin operator, $\hat{\mathbf{S}}^2$, as a penalty term to the original molecular Hamiltonian. We verify that this approach provides good convergence behavior towards the lowest energy eigenstate using direct matrix exponentiation for Imaginary Time Evolution (ITE). Numerical simulations were performed on the H$_2$ molecule and a square tetrahydrogen cluster using measurement-assisted unitary approximation in QITE. QITE demonstrates faster convergence to the ground state with broken symmetry (BS) compared to HF, particularly after the molecule exhibits a diradical character of 0.56 for H$_2$. Prior to this point, HF remains more effective, suggesting a transition threshold of diradical character for wave function selection. Additionally, the overlap analysis with Complete Active Space Configuration Interaction (CAS\text{-}CI) wave function shows that BS has a larger initial overlap than HF in higher-spin, multi-configurational systems like triple bond dissociation in N$_2$. This method provides a pathway for improved energy simulations in open shell systems, where wave function accuracy significantly impacts downstream quantum algorithms and practical applications in quantum chemistry.
\end{abstract}


\section{Introduction}
Ground state energy estimation (GSEE) is the problem of finding the lowest energy eigenstate of a given Hamiltonian, and is considered to a promising application of quantum computers that may show the first realization of practical quantum advantage~\cite{r24Aspuru-Guzik2005}. This problem is of utmost importance for drug discovery~\cite{r12Sala2025}, semiconductor and battery manufacturers~\cite{r13Kim2022}, and among chemists and physicists worldwide.

Various quantum computing approaches like Quantum Phase Estimation~\cite{r24Aspuru-Guzik2005},~\cite{r22Kitaev1995},~\cite{r23Abrams1999}, Variational Quantum Eigensolver (VQE)~\cite{r14Belaloui2024}, Quantum Monte Carlo (QMC)~\cite{r44Huggins2022}, Adiabatic State Preparation (ASP)~\cite{r24Aspuru-Guzik2005}, and QITE~\cite{r4Motta2020} are proposed to solve this common problem of GSEE. Despite lots of improvements to date, the application of these quantum algorithms to strongly correlated systems, such as multinuclear transition metal complexes and molecules under covalent bond dissociation, remains a challenge~\cite{r1Sugisaki2016},~\cite{r25Sugisaki2019},~\cite{r26lee2023}. Under the strong correlation, the ground state wave function shows strong multi-configurational character, and the HF wave function is not a good approximation of the ground state. The small energy gap between the ground and excited states is another important feature of the strong correlation, and ASP and QITE usually suffer from slow convergence. Acceleration of the GSEE through improving the convergence behavior is an important issue. 
Recently, we reported that the ASP-based GSEE for the strongly correlated systems can be accelerated by using broken-symmetry wave function as the initial wave function, in conjunction with adding an electron spin $\hat{\mathbf{S}}^2$ operator to the Hamiltonian as a penalty~\cite{r3Sugisaki2022}. In this work, we report the improvement of the convergence behavior in QITE for strongly correlated systems by adopting the broken-symmetry wave function.

\subsection{Open shell systems}
Open shell systems are those molecular systems carrying unpaired electrons. When the molecule has more than one unpaired electron, electronic states of different spin multiplicities appear in the low energy region, and they are representative of strongly correlated systems.  Open shell systems play an important role in chemistry and materials science as potential candidates for magnetic materials~\cite{r27Coronado2020} and nonlinear optical materials~\cite{r28nakano2015}.  In addition, multinuclear transition metal complexes containing unpaired d-electrons are involved in a number of catalytic processes and enzymes, such as Mn$_4$CaO$_5$ cluster in the photosystem II~\cite{r29umena2011} and FeMoco cluster in nitrogenase~\cite{r30spatzal2016}. It is important to note that open shell electronic structure appears even in small, simple molecules such as H$_2$ when covalent bond dissociation takes place.

Reactants involved in a chemical reaction are in the state of equilibrium until the reaction initiates. As the reaction proceeds, bond breaking takes place within the reactants before they form new bonds. As the interatomic distance increases, the energy gap between the ground state and the first excited state decreases. The study of these reaction pathways and intermediate stages during a chemical reaction is important as it helps in designing proper conditions for the reaction, and predicting the stability of molecules and radicals formed during the entire reaction. Sugisaki et al. studies the quasi-reaction pathway of a beryllium (Be) atom insertion into a hydrogen (H\textsubscript{2}) molecule~\cite{r10Sugisaki2022}. This reaction is an example of chemical reactions which encounter avoided crossings between the ground state and the first excited state in the Potential Energy Surface (PES). 


\subsection{Current state and challenges in quantum computation for GSEE of open shell systems}

The conventional practice for GSEE in the noisy intermediate-scale quantum (NISQ) era~\cite{preskill2018nisq} is to use VQE with the Unitary Coupled Cluster (UCC) ansatz with the spin-restricted HF (RHF) wave function as the reference~\cite{r31anand2022}.
However, RHF is a poor approximation for open shell systems with antiferromagnetic couplings between unpaired electrons.
Consequently, consideration of higher-order excitation operators is necessary to obtain an accurate ground-state energy in VQE-UCC.
In the real- and imaginary-time evolution-based ASP and QITE, the convergence speed is strongly controlled by the energy gap between the ground and the excited states.
Narrowing energy gap results in decreasing convergence rate in the GSEE of open shell systems.

Several approaches have been proposed to tackle this problem.
Instead of single configurational HF wave functions, techniques using multi-configurational wave functions such as Complete Active Space Self-Consistent Field (CASSCF) with smaller active space have been well studied.
For example, multireference UCC methods like Multireference-Unitary Coupled Cluster with partially Generalized Singles and Doubles (MR-UCCpGSD) ansatz~\cite{r10Sugisaki2022} were proposed for VQE.
An approach using the CASSCF wave function as the initial wave function has also been investigated in ASP~\cite{r32kremenetski2021}.
These methods greatly improve the convergence, but solving CASSCF itself becomes intractable when a large active space is required to capture strong correlations.
The need for deep quantum circuits to encode multiconfigurational reference wave functions is another problem in quantum computation. It is highly desirable to develop theoretical methods that are applicable to strongly correlated systems, with shallow quantum circuits for initial state preparation.

We address the problem of slow convergence of QITE for molecules by proposing a new variant that takes into account the open shell character of that molecule throughout its PES during  bond dissociation. We have figured out that the introduction of spin penalty to the Hamiltonian during QITE, started with BS wave function, provides an improvement to the time it takes to reach the ground state.
We demonstrate our approach to matrix exponentiation for the time evolution of BS wave function with ITE on the case of N$_2$ molecule.
As a proof of concept, to show that it performs well with unitary approximation as in QITE, we perform numerical emulations for the H$_2$ molecule PES and square tetrahydrogen (P4) cluster~\cite{r37Paldus1993} with interatomic distances being 2.0 Bohr. We identify the region in PES of H$_2$ and N$_2$ molecules where our method offers a significant speedup in the convergence rate. The region is identified using the diradical character $y$ as a measure of open shell electronic configuration, which we use to select the appropriate starting wave function for QITE.

\section{Background}
\label{sec:background}

\subsection{Imaginary-time Schr\"{o}dinger equation}
Imaginary time is a mathematical concept used in statistical mechanics, quantum field theory, and cosmology.
It involves rotating the time axis in complex space~\cite{r43Wick1954}, from real time $t$ to imaginary time $\tau$ as $t\rightarrow i\tau$.
Exponential integrals such as $e^{i\hat{H}t}$ oscillate, making it often difficult to evaluate their long-term properties.
Rotating to imaginary time maps the oscillations to a smooth function with decay rate dependent on the size of the integral components.
Thus, performing ITE allows the function to decay towards the lowest energy eigen state of the system evolving in time.

The imaginary-time Schr\"{o}dinger equation takes the form
\begin{equation}
    -\partial_{\tau} |\Phi(\tau)\rangle = \hat{H}|\Phi(\tau)\rangle.
    \label{eq:qite_schroedinger}
\end{equation}
The ground state $|\Psi_0\rangle$ of the Hamiltonian $\hat{H}$, with $\langle\Phi(0)|\Psi_0\rangle\neq0$, is given by the long-term limit of~(\ref{eq:qite_schroedinger}),
\begin{equation}
    |\Psi_0\rangle = \lim_{\tau \to \infty} \frac{|\Phi(\tau)\rangle}{\| |\Phi(\tau)\rangle \|}.
    \label{eq:qite_ground_state}
\end{equation}
QITE decomposes the Hamiltonian into $\hat{H} = \sum_m \hat{h}[m]$, where all $\hat{h}[m]$ are at most $k$-local, and Trotterizes the generated evolution,
\begin{equation}
    e^{-\tau\hat{H}} = (e^{-\Delta\tau\hat{h}[1]}e^{-\Delta\tau\hat{h}[2]}...)^{\tau/\Delta\tau} + \mathcal{O}(\Delta\tau).
    \label{eq:qite_evolution}
\end{equation}

Various methods exist to approximate the non-unitary evolution of~(\ref{eq:qite_evolution}) by a unitary operator~\cite{r17Liu2020ProbabilisticNG},~\cite{r18Ammar2017},~\cite{r19Gily2019}.
Our approach relies on the measurement-assisted unitary circuits introduced in~\cite{r22Kitaev1995}.
The Trotterized operators $e^{-\Delta\tau\hat{h}[m]}$ acting on $k$ qubits can be approximated by a unitary operator $e^{-i\Delta\tau\hat{A}[m]}$ acting on $D\geq k$ qubits.
The $D$-local operators $\hat{A}[m]$ can be determined following the procedure in~\cite{r4Motta2020}.
The domain size $D$ grows with the correlation length in the system.
For an $N$-qubit Hamiltonian and bounded correlation length, the cost of a single QITE step is linear in $N$ in space, and polynomial in $N$ in time~\cite{r4Motta2020}, offering an exponential speedup compared to classical ITE.

This method of QITE for GSEE can be implemented without the need for deep circuits and ancillae.
However, with the existing approach only, the performance of this algorithm is not very impressive when applied to systems with high correlation, needing a larger domain size and long evolution time. It is certain that the decaying wave function will converge to the minimum ground state at some point. We have exploited the advantage of a BS wave function, implementing a spin penalty term-inspired Hamiltonian such that the same algorithm converges to the ground state a lot faster than it used to before. We performed numerical emulation for finding the ground state energy of a H\textsubscript{2} molecule during H--H bond dissociation and a P4 cluster during 2H--2H bond dissociation. This serves as a testament that the BS wave function performs well for the QITE algorithm with proper unitary approximation.

\subsection{Spin operators}

Spin is an intrinsic angular momentum without a classical analog, and it is described by a spin operator $\hat{\mathbf{S}} = (\hat{S}_x, \hat{S}_y, \hat{S}_z)$.
The spin operator components obey the usual angular momentum commutation relations $[\hat{S}_j,\hat{S_k}] = i\epsilon_{jkl}\hat{S}_l$, where $\epsilon_{jkl}$ is the Levi-Civita symbol, $j,k,l=x,y,z$, and we work in units where $\hbar=1$.
The spin magnitude is described by the square of the spin operator $\hat{\mathbf{S}}^2$, which commutes with components of the spin operator $[\hat{\mathbf{S}}^2,\hat{S}_i]=0$, for $i=x,y,z$.
This leads to a common set of eigenvectors for $\hat{\mathbf{S}}^2$ and one of the spin components, usually taken to be $\hat{S}_z$. These eigenvectors $|s,m\rangle$ are labeled by the spin $s$ and its projection along the z-direction $m$, and satisfy the following eigenvalue equations,
\begin{align}
    \hat{\mathbf{S}}^2 |s,m\rangle & = s(s+1)|s,m\rangle, & s & = 0,\frac{1}{2},1,\frac{3}{2},\ldots \label{eq:eigenvalue1}\\
    \hat{S}_z |s,m\rangle & = m |s,m\rangle, & m & = -s, -s+1, \ldots, s \label{eq:eigenvalues2}.
\end{align}
The operator $\hat{\mathbf{S}}^2$ can be written in the following useful form,
\begin{equation}
    \hat{\mathbf{S}}^2 = \frac{1}{2} \left( \hat{S}_+\hat{S}_- + \hat{S}_-\hat{S}_+ \right) + \hat{S}_z^2,
    \label{eq:S_squared_single}
\end{equation}
where $\hat{S}_{\pm}=\hat{S}_x\pm i\hat{S}_y$ are the raising/lower operators.

As an example, consider a single electron.
Its spin component operators are given by the Pauli matrices $\hat{S}_i=\frac{1}{2}\hat{\sigma}_i$, for $i=x,y,z$.
Since $s=\frac{1}{2}$ for an electron, there are two possible states,
\begin{equation}
    |\alpha\rangle = \left|\frac{1}{2},+\frac{1}{2}\right\rangle, \quad\text{and}\quad |\beta\rangle = \left|\frac{1}{2},-\frac{1}{2}\right\rangle,
    \label{eq:single_spin}
\end{equation}
where we introduce $\alpha$ and $\beta$ as a convenient notation for the spin-up and spin-down states of a single electron.

Let us consider two systems with spin operators $\hat{\mathbf{S}}_1$ and $\hat{\mathbf{S}}_2$.
The total spin operator of the combined system is $\hat{\mathbf{S}} = \hat{\mathbf{S}}_1 + \hat{\mathbf{S}}_2$, with possible spin values $S$ and corresponding z-component $M$ given by the addition theorem for angular momenta,
\begin{align}
    S & = s_1+s_2, s_1+s_2-1, \ldots, |s_1 - s_2|, \label{eq:eigenvalue_total}\\
    M & = -S, -S+1, \ldots, S.
    \label{eq:z_component}
\end{align}
Here, $s_1$ and $s_2$ are the spin values of the individual subsystems, and we use capitalized $S$ to emphasize the spin value for the combined system.
Equation~(\ref{eq:S_squared_single}) can be generalized to the case of two subsystems,
\begin{equation}
    \hat{\mathbf{S}}^2 = \hat{\mathbf{S}}_1^2 + \hat{\mathbf{S}}_2^2 + \hat{S}_{1+}\hat{S}_{2-} + \hat{S}_{1-}\hat{S}_{2+} + 2\hat{S}_{1z}\hat{S}_{2z}.
    \label{eq:S_squared_expanded}
\end{equation}
The total $N$-electron spin eigenvectors are denoted by $|N,S,M\rangle$ and they satisfy similar eigenvalue equations to~(\ref{eq:eigenvalue1}) and ~(\ref{eq:eigenvalues2}), provided the total spin value satisfies the constraint of~(\ref{eq:eigenvalue_total}).

\subsection{The broken-symmetry wave function}

The electronic configuration of the highest spin state with spin quantum number $S=N/2$ with spin magnetic quantum number $|M|=S$ can be described by a single determinant.
On the other hand, low-spin ($S < N/2$) states and $|M| \ne S$ sublevels of the highest spin state show inherent multi-determinant character.
This prevents the application of conventional single-determinant methods such as Density Functional Theory (DFT) to compute open shell low-spin states.
BS wave function addresses this issue by using a linear combination of the high- and low-spin wave functions to generate a spin-mixed single configurational wave function,
\begin{equation}
    |\text{BS}(N,M)\rangle = \sum_{j=0}^{N/2} a_j |N,S=j,M\rangle,
    \label{eq:generalized_bs_wf}
\end{equation}
where the probability amplitudes $a_j$ can be obtained by following the branching procedure explained in Appendix~\ref{sec:appA}.
BS wave functions have been widely used in spin chemistry to calculate exchange interactions between unpaired electrons that determine the energy gap between spin states~\cite{r33YAMAGUCHI1975,r34shoji2006}. On the quantum computing side, the BS wavefunction is described by a single electron configuration, which has the advantage that state preparation can be done with Pauli-X gates only.

Let us consider two spin eigenfunctions $|2,0,0\rangle$ and $|2,1,0\rangle$. Taking their equal superposition generates the spin state $|\alpha\beta\rangle$ as follows,
\begin{align}
    &\quad |\alpha\beta\rangle = \frac{1}{\sqrt{2}}|2,0,0\rangle + \frac{1}{\sqrt{2}}|2,1,0\rangle.
    \label{eq:BS_H2}
\end{align}
This wave function corresponds to a BS state of the H\textsubscript{2} molecule.
The notation $|\alpha\beta\rangle$ in~(\ref{eq:BS_H2}) does not represent spin-up and spin-down state of two electrons.
Rather, it denotes the occupancy of active molecular orbitals.
The BS state in~(\ref{eq:BS_H2}) describes a state where the lower orbital is occupied by a single spin-up electron and the second orbital contains a single spin-down electron.
More generally, a molecule with $k$ active orbitals is denoted by $|c_1c_2\ldots c_k\rangle$, where $c_i\in\{2,\alpha,\beta,0\}$.
Occupancy $c_i=2$ means the orbital contains paired electrons, while $c_i=0$ is an unoccupied orbital.

\begin{figure}[t]
    \centering
    \begin{subfigure}{0.6\columnwidth}
        \centering
        \includegraphics[width=\linewidth]{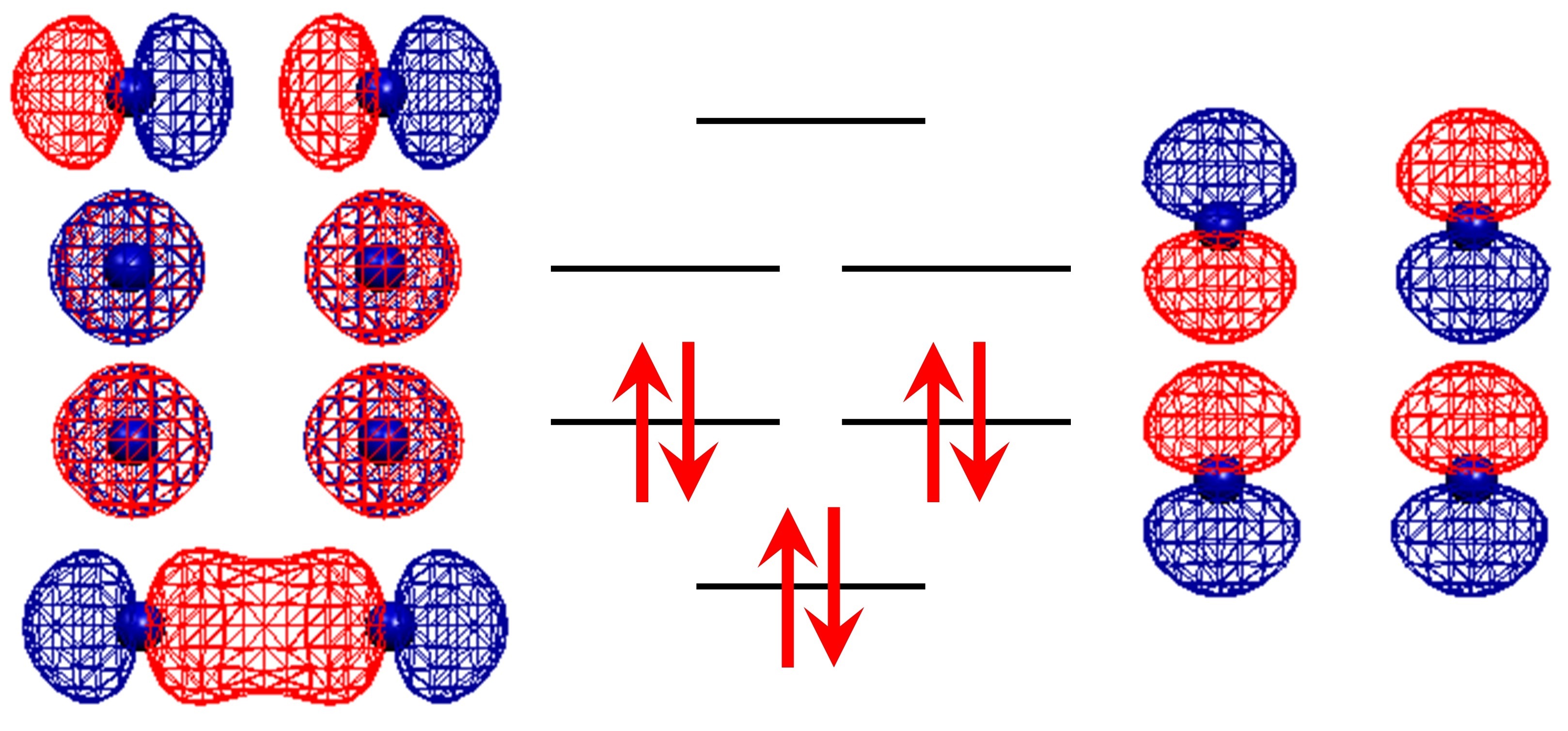}
        \caption{}
        \label{fig:subfig1a}
    \end{subfigure}
    \hfill
    \begin{subfigure}{0.6\columnwidth}
        \centering
        \includegraphics[width=\linewidth]{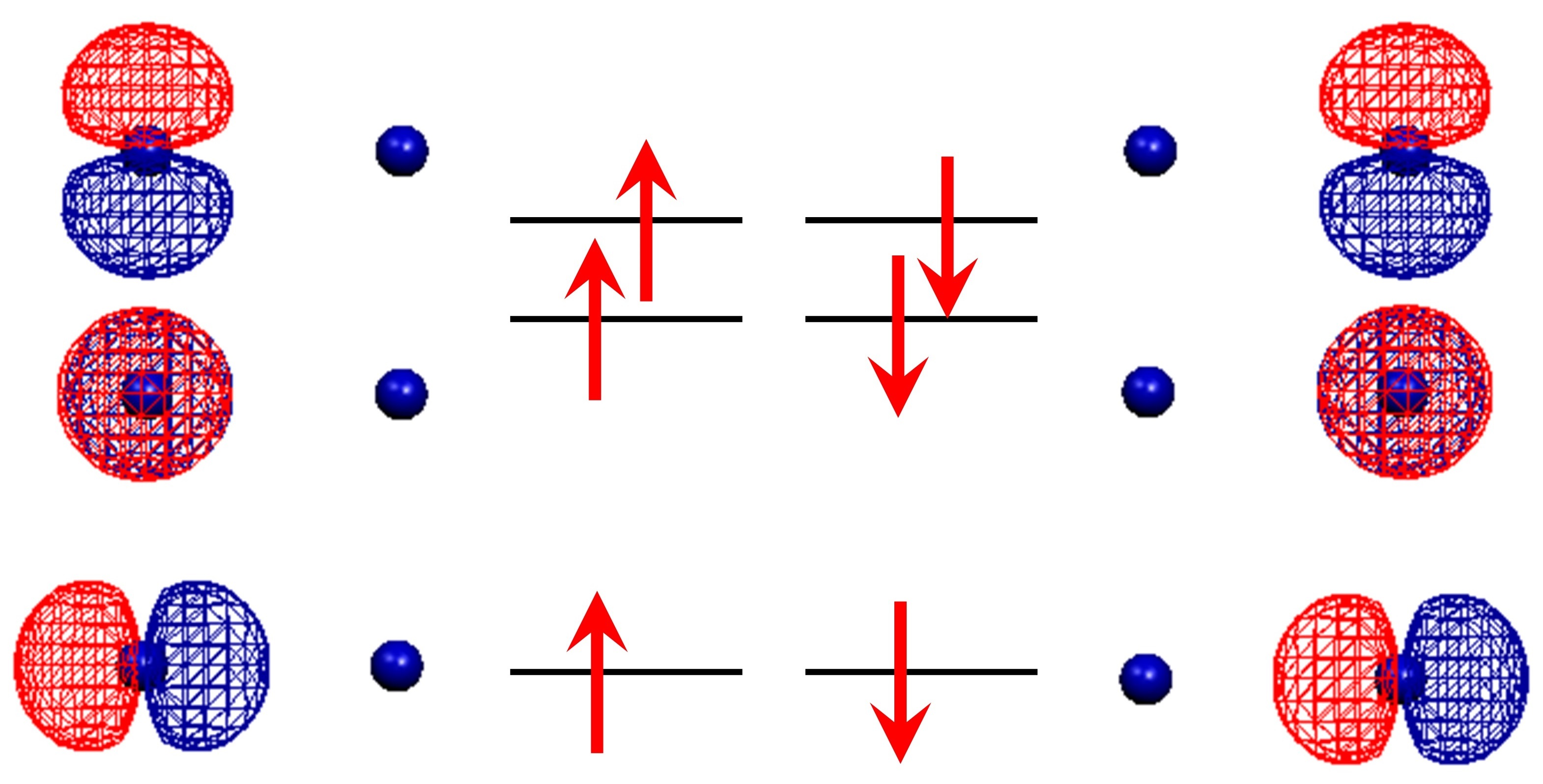}
        \caption{}
        \label{fig:subfig1b}
    \end{subfigure}
    \caption{Active orbitals of the N\textsubscript{2} molecule.  The blue dots represent N atoms. (a) RHF canonical orbitals. Arrows specify the RHF electronic configuration. (b) Localized orbitals constructed from the natural orbitals computed at the BS-UHF level. Arrows specify the electron occupancies of the BS3.}
    \label{fig:n2_geometries}
\end{figure}

Let us consider a N\textsubscript{2} molecule which has triple bonds (two $\pi$ bonds and a $\sigma$ bond) between two N atoms. We represent the BS wave function corresponding to two $\pi$-bond dissociation as BS2 and both $\pi$ and $\sigma$ bond dissociation as BS3. Fig.~\ref{fig:n2_geometries} compares the electronic configuration of RHF and BS3 active orbitals.
In the active space consisting of valence $\sigma$ and $\pi$ orbitals, the electron configuration of the BS2 and BS3 are represented as $|2\alpha\alpha\beta\beta0\rangle$ and $|\alpha\alpha\alpha\beta\beta\beta\rangle$, respectively. BS2 and BS3 wave functions are described as the sum of states of different spin multiplicities with zero spin magnetic quantum number, 
\begin{align}
    |2\alpha\alpha\beta\beta0\rangle & = \frac{1}{\sqrt{6}}|4,2,0\rangle +\frac{1}{\sqrt{2}}|4,1,0\rangle + \frac{1}{\sqrt{3}}|4,0,0\rangle , 
    \label{eq:BS2_N2} \\
    |\alpha\alpha\alpha\beta\beta\beta\rangle & = \frac{1}{\sqrt{20}}|6,3,0\rangle +\frac{1}{{2}}|6,2,0\rangle + \frac{3}{\sqrt{20}}|6,1,0\rangle\nonumber \\
    & + \frac{1}{{2}}|6,0,0\rangle. 
    \label{eq:BS3_N2}
\end{align}
Again, the probability amplitudes are obtained from the branching procedure in Appendix~\ref{sec:appA}.

\section{QITE with Penalty and Broken-Symmetry Wave Functions}

In order to avoid the problem of narrowing energy gap, we introduce a penalty term to the original Hamiltonian $\hat{H}$,
\begin{align}
    \hat{H}' = \hat{H} + c\hat{\mathbf{S}}^2,
    \label{eq:Hamiltonian_with_penalty}
\end{align}
where $c$ controls the strength of the penalty term.
A similar approach was applied to speed up convergence to the ground state in the context of ASP in~\cite{r3Sugisaki2022}.

The addition of the penalty term adds an energy contribution proportional to $S(S+1)$ for each spin component present in the wave function.
Low spins suffer a small penalty in comparison to large spins, leading to an increase in the energy gap between the singlet ground state and the excited states belonging to different spin multiplicities.
Ultimately, this leads to faster convergence to the ground state.

\begin{figure}[t]
    \centering
    \begin{subfigure}{0.48\columnwidth}
        \centering
        \includegraphics[width=\linewidth]{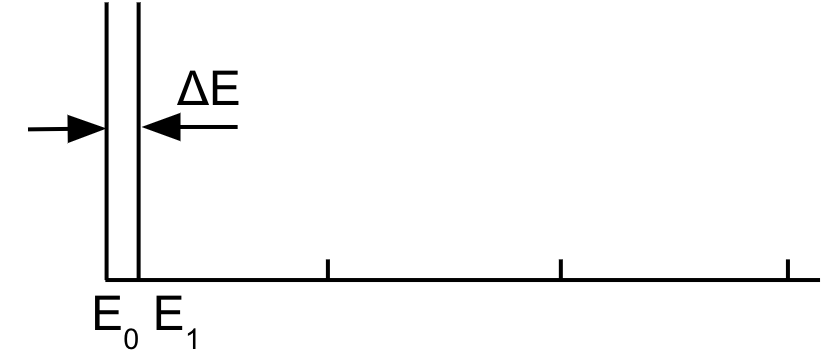}
        \caption{}
        \label{fig:subfig2a}
    \end{subfigure}
    \hfill
    \begin{subfigure}{0.48\columnwidth}
        \centering
        \includegraphics[width=\linewidth]{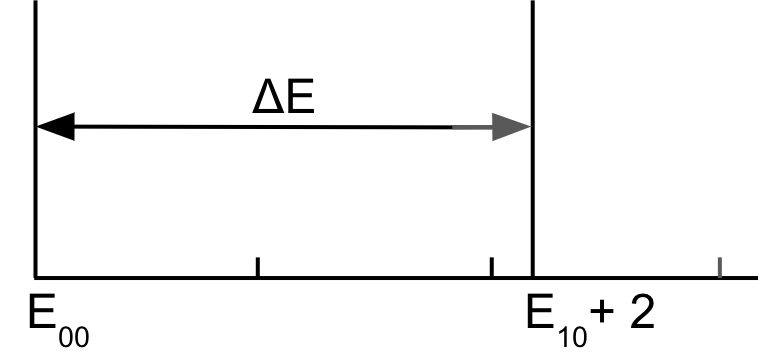}
        \caption{}
        \label{fig:subfig2b}
    \end{subfigure}
    \caption{(a) Energy gap between the singlet ground state and singlet excited state corresponding to the delocalized HF wave function. (b) Energy gap between the singlet ground state and the excited triplet state corresponding to the localized BS wave function after introducing penalty.}
    \label{fig:h2_energy_gaps}
\end{figure} 

It is important to note that the HF wave function is the eigenfunction of the $\hat{\mathbf{S}}^2$ operator, and in the bond dissociation limit of the H$_2$ molecule, the HF wave function is expressed as a superposition of the singlet ground state and the singlet excited state. Since the excited state of the same spin multiplicity as the ground state contributes to the wave function, it is difficult to apply a penalty term to lift the quasi-degeneracy.  
On the contrary, the BS wave function for H$_2$ molecule in the bond dissociation limit is described as the linear combination of the singlet ground state and the triplet excited state wave functions. In this case, introduction of the $\hat{\mathbf{S}}^2$ operator to the Hamiltonian raises the energy of the triplet state lifting the  quasi-degeneracy of the two states.

Fig.~\ref{fig:subfig2a} shows the schematic view of the energy gap ($E_1 - E_0$) between the singlet ground state and singlet excited state in the H$_2$ molecule. In the vicinity of the equilibrium geometry, the energy gap is large, but the gap becomes smaller when the covalent bond dissociation takes place. In the bond dissociation limit, the ground-state wave function is well approximated by the linear combination of the RHF configuration and two-electron excited configuration from the RHF state, $|\Psi_0\rangle = \frac{1}{\sqrt{2}} (|20\rangle - |02\rangle)$.

This wave function is expressed as a linear combination of the singlet ground state and the singlet excited state. In contrast, when the molecular orbitals localized onto spin sites are used as the wave function expansion, the ground state wave function is approximated by a linear combination of open shell configurations: $|\Psi_0\rangle = \frac{1}{\sqrt{2}} (|\alpha\beta\rangle + |\beta\alpha\rangle)$. Note that the molecular orbitals localized to a spin site can be constructed by mixing the highest occupied molecular orbital (HOMO) and the lowest unoccupied molecular orbital (LUMO) of the RHF canonical orbitals.
The BS wave function $|\alpha\beta\rangle$ is described by a linear combination of the singlet ground state and the triplet excited state. The energy gap between two states becomes small when the bond dissociation occurs, but introduction of the spin penalty term $\hat{\mathbf{S}}^2$ to the Hamiltonian can enlarge the gap, as shown in Fig.~\ref{fig:subfig2b}.   

\section{Computational conditions of QITE}
\label{sec:computational conditions}

We implemented QITE algorithm and demonstrated the advantage of using BS wave function as the initial wave function with penalty introduced in the Fermionic Hamiltonian using exact classical emulation. RHF and BS-UHF calculations were carried out using \texttt{GAMESS-US} software~\cite{r35Zahariev2023}. One and two electron atomic orbital integrals computed with \texttt{GAMESS-US} were transformed to molecular orbital integrals using our-own Python3 code, and the Fermionic Hamiltonian was constructed with \texttt{OpenFermion}~\cite{r36McClean_2020} package.

We focused on three molecular systems: H\textsubscript{2} molecule, N$_2$ molecule, and P4 cluster. The study of bond dissociation of a N$_2$ molecule in this work is limited to ITE with matrix exponentiation only, to verify that our work also achieves advantage for larger molecules. For N$_2$ molecule, STO-3G basis set was used with six electrons in twelve spin orbitals. Statevector based numerical simulations were performed for the H$_2$ molecule and a P4 cluster with measurement-assisted unitary approximation for QITE. In the study of the ground state energy of the H\textsubscript{2} molecule under a single bond dissociation, we used STO-6G basis set with two electrons in four spin orbitals. In case of a P4 cluster, STO-3G basis set was used with four electrons in eight spin orbitals for the dissociation of intermolecular hydrogen bonds. To perform Fermion-to-qubit mappings, Symmetry-Conserving Bravyi--Kitaev Transformation (SCBKT) was used [8]. 1500 total iteration steps were taken for the iterative calculation of $\hat{A}[m]$ for H\textsubscript{2} and P4 cluster with a step size of 0.01. The domain sizes for H\textsubscript{2} and P4 cluster were taken to be 2 and 6 respectively.
The strength of the penalty term in~(\ref{eq:Hamiltonian_with_penalty}) was set to $c=1$.

\section{Results and discussion}

\subsection{N$_2$ using ITE by matrix exponentiation}
We have investigated the time of convergence to the ground state and the fidelity against CAS-CI energy for N\textsubscript{2} molecule under different bond lengths. Simulations were carried out for different ranges of bond length of N--N atoms, RHF (1 Å - 2.9 Å), BS2 (1.6 Å - 3 Å), BS3 (2.1 Å - 3 Å). BS3 converged to BS2 when the bond length was shorter than 2.1 Å, and BS2 calculations for the N$_2$ molecule with bond length shorter than 1.6 Å converged to the RHF solution, since N$_2$ molecule exhibits very small open shell character in the range below them.
We define the convergence time as the time required to achieve chemical precision, when the energy difference between the ITE state and the CAS-CI state is $\Delta E = E_{\text{ITE}} - E_{\text{CAS-CI}} \leq1.0\ \text{kcal}\;\text{mol}^{-1}$.

\begin{figure*}[t]
    \centering
    \includegraphics[width=0.9\textwidth]{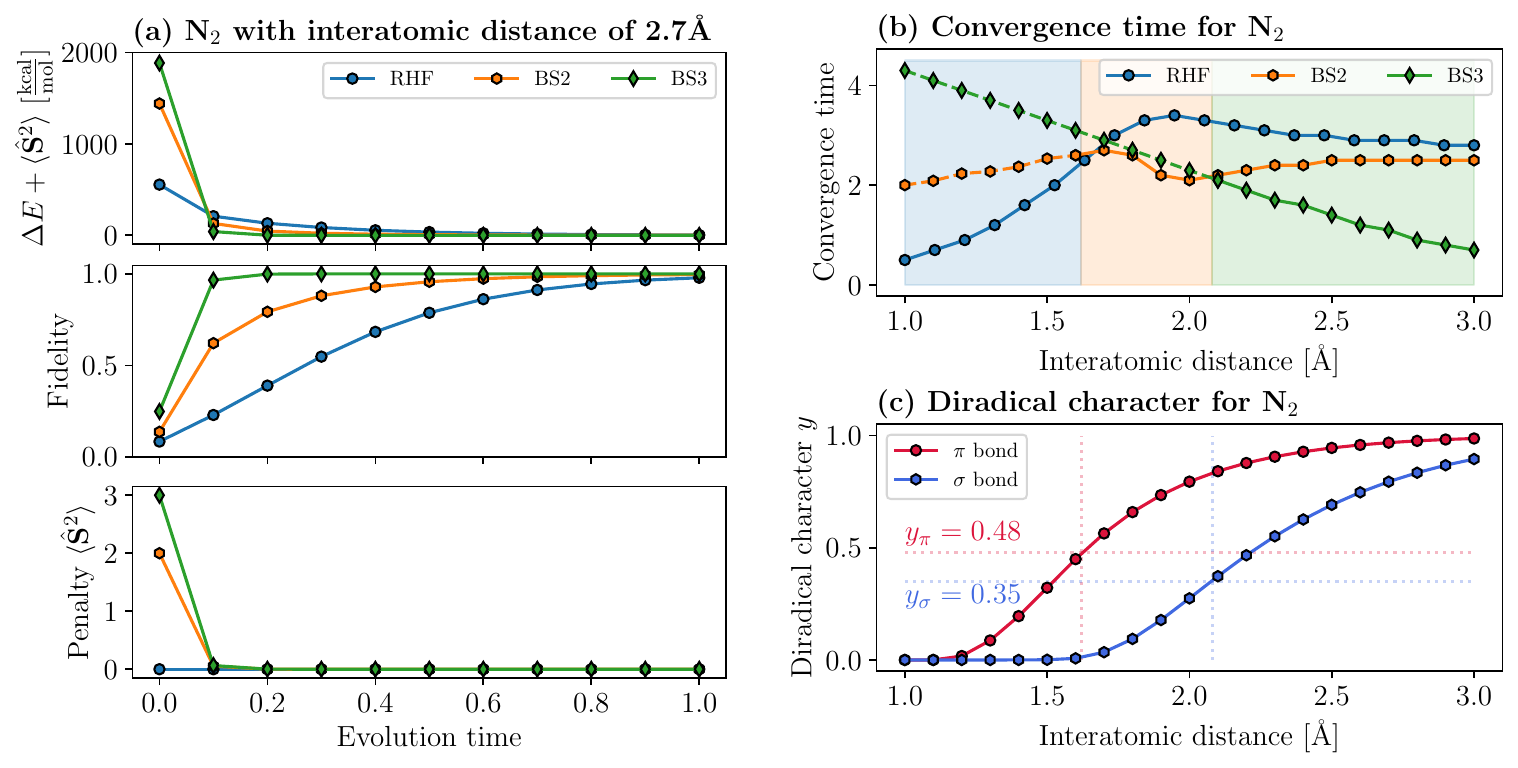}
    \caption{(a) Comparison of the convergence behavior of ITE in N$_2$ molecule with the RHF, BS2, and BS3 wave functions as the starting wave function. Top panel shows the energy difference $\Delta E$ during the evolution. Fidelity of ITE evolved wave function with respect to the CAS-CI wave function in the middle panel, and contribution of the penalty term $\langle \hat{\mathbf{S}}^2 \rangle$ in the bottom panel. (b) Time of convergence to attain the chemical precision $\Delta E \le$ 1.0 kcal mol$^{-1}$ for bond length from 1.0 Å to 3.0 Å. Background colors show the recommended region of wave function as starting wave function depending on the N--N bond length. (c) Diradical character $y$ at the region of bond dissociation. The points of intersection of red and blue lines with corresponding red and blue curves indicate the recommended points to change the initial wave functions described in (b).}
    \label{fig:n2_matrix_exponentiation_results}
\end{figure*}

Fig.~\ref{fig:n2_matrix_exponentiation_results}a shows faster convergence towards the ground state energy for BS2 and BS3 than RHF inspired ITE for a N--N bond length of 2.7 Å.
In addition to this, an important observation to note is, that the initial fidelity of both the BS2 (\(\approx0.14 \)) and BS3 (\(\approx 0.25 \)) wave functions with respect to the CAS-CI wave function at 2.7 Å bond length is larger than that of RHF wave function (\(\approx0.08 \)).
The initial fidelity of the BS2 and BS3 wave functions in the bond dissociation limit is \(\approx0.12 \) and \(\approx0.25 \) respectively.
The contribution of the penalty term in~(\ref{eq:Hamiltonian_with_penalty}) is shown in the lower panel of Fig.~\ref{fig:n2_matrix_exponentiation_results}a.
For the RHF wave function, $\langle\hat{\mathbf{S}}^2\rangle$ remains zero throughout the evolution, resulting in slow convergence towards the ground state.
Interestingly, when the initial state is a BS wave function, the penalty contributes significantly only in the early stage of the evolution.
This brief but large contribution ensures the initial state quickly evolves towards the ground state.

Next, we investigate whether the BS wave function is a suitable initial state for all bond lengths.
Fig.~\ref{fig:n2_matrix_exponentiation_results}b shows that this is not always the case.
For shorter bond lengths (blue region), the RHF converges faster than both BS2 and BS3.
As the bond length increases, the convergence time for BS wave functions decreases, while the performance of the RHF wave function worsens.
For intermediate bond lengths (yellow region), the BS wave functions outperform the RHF wave function, with BS2 performing slightly better than BS3.
Finally, we observe that for long bond lengths (green region), the BS3 wave function is far more suitable initial state than RHF or even BS2. It is also important to note here that we used the knowledge of convergence time versus interatomic distance for the range described above for the case of BS2 and BS3 to extrapolate the plots for bond lengths ranging from 1.0 Å to 3.0 Å in Fig.~\ref{fig:n2_matrix_exponentiation_results}b.

In order to determine when a particular wave function is most suitable as the initial state, we compute the diradical character as a function of bond length.
Diradical character is a measure of the open shell character of the wave function, and it can be calculated at the spin-projected spin-unrestricted HF (PUHF) level~\cite{Doehnert1980},
\begin{align}
    y_i^{\mathrm{PUHF}} = 1 - \frac{2(1 - n_{\mathrm{LUNO}+i})}{1 + (1 - n_{\mathrm{LUNO}+i})^2}.
    \label{eq:diradical_character}
\end{align}
Here, $n_{\mathrm{LUNO}+i}$ is the occupation number of the $i$-th lowest unoccupied natural orbital (LUNO$+i$).
For convenience, we denote the diradical character $y^{\text{PUHF}}_i$ as $y$ throughout the rest of the paper.
It ranges from $0$ to $1$, a molecule with $y>0$ is considered to have open shell character which becomes more prominent as $y$ increases.
At the extremes, $y=0$ for a closed shell molecule, and $y=1$ for a pure diradical.

For N\textsubscript{2} bond dissociation, we focus on two diradical characters $y_\pi$ and $y_\sigma$ for open shell characters of valence $\pi$ and $\sigma$ bonds, respectively, as shown in Fig.~\ref{fig:n2_matrix_exponentiation_results}c.
Numerical simulations reveal that ITE with BS2 converges as fast as the conventional RHF-reference ITE within the bond length of 1.62 Å corresponding to the diradical character $y_\pi$ = 0.48. For further dissociation, BS2 wave function performs better until the bond length reaches 2.08 Å with $y_\sigma$ = 0.35. Beyond this bond length, BS3 is the most suitable to use as a starting wave function.

However, it remains an open problem to use the diradical character to determine the proper starting wave function without the aid of these simulation works.

\subsection{H$_2$ using QITE }
We now focus on the H$_2$ molecule and get into the details of the behavior of QITE throughout its PES.
As discussed earlier, the maximum spin multiplicity involved in the BS wave function for a H$_2$ molecule is $S$ = 1.
This leads to an extra penalty contribution of $S(S+1)=2$ to the energy of the excited triplet, leading to an increase in the energy gap with respect to the ground state.

\begin{figure*}[t]
    \centering
    \includegraphics[width=0.9\textwidth]{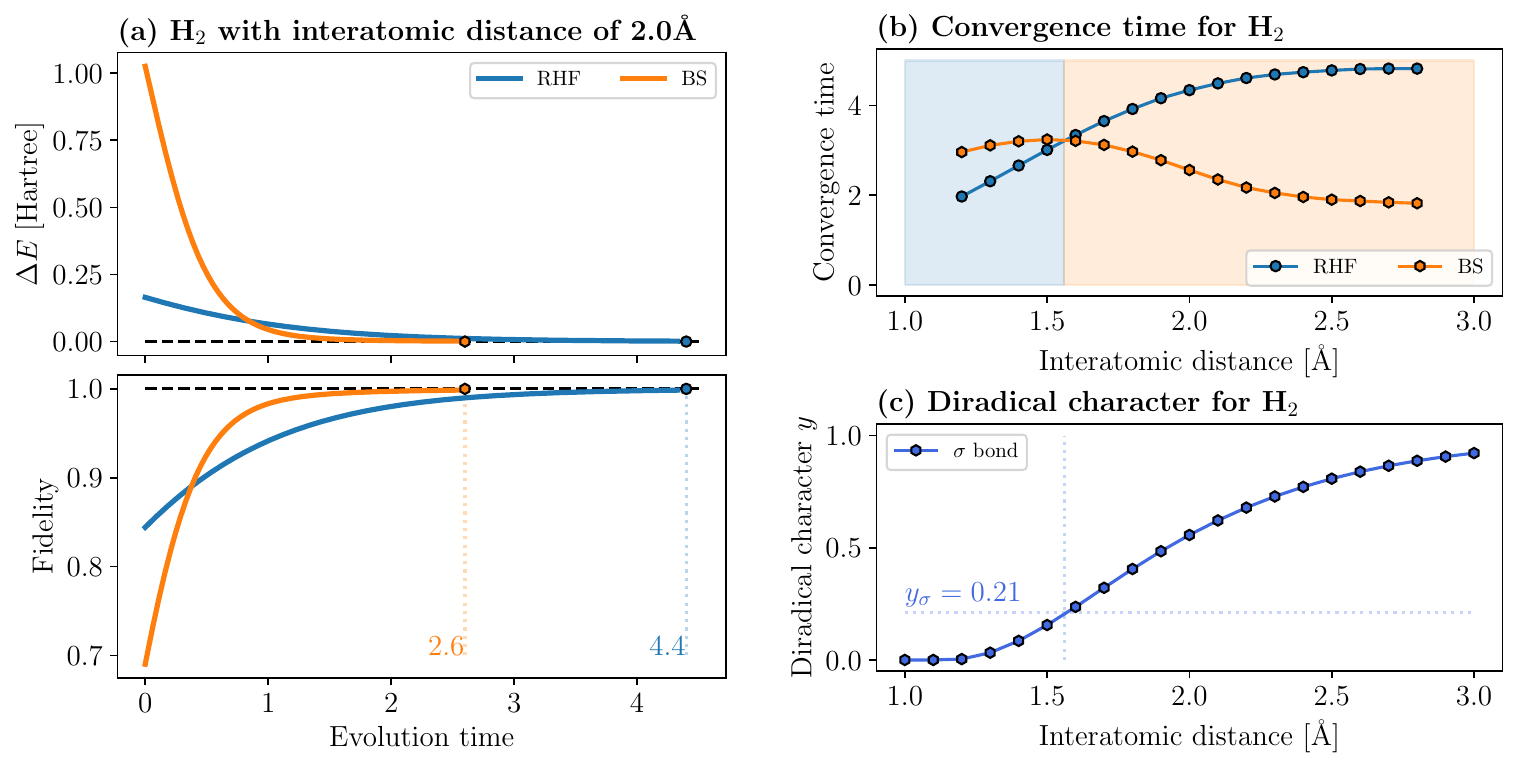}
    \caption{(a) Comparison of the convergence behavior of QITE in H$_2$ molecule with the RHF and BS wave functions as the starting wave function. Top panel shows the energy difference between the ground state energy calculated using CAS-CI and QITE with unitary approximation. Dashed line represents chemical precision given by $1.59\times10^{-3}$ Hartree. Bottom panel shows fidelity of QITE evolved wave function with respect to the CAS-CI wave function. Dashed line marks unit fidelity. (b) Recommended region of wave function as starting wave function depending on the bond length between hydrogen atoms. (c) Diradical character $y$ at the region of bond dissociation. The point of intersection of two straight lines with the curve indicate the recommended point to change the initial wave function described in (b).}
    \label{fig:h2_simulation_results}
\end{figure*}

Fig.~\ref{fig:h2_simulation_results}a shows the difference in ground state energy with two different wave function driven QITE with CAS-CI values and the comparison of fidelities. The chemical precision could be attained within 260 steps using BS wave function, while it took 440 steps with HF wave function for a bond length of 2.0 Å with diradical character $y = 0.56$. This simulation was carried out with domain size of 2 considering two electrons in the Hamiltonian. Here we discard the concept of locality while approximating the unitary operator from a nonunitary time evolution. 

Similar to what we observed during matrix exponentiation of the N$_2$ molecule, it is not wise to use a BS wave function in all the context of H$_2$ molecule too. During an equilibrium, at bond length of around 0.7 Å and upto 1.56 Å, the HF wave function is a better choice for starting wave function. Bond length of 1.56 Å corresponds to the diradical character $y = 0.21$. For H\textsubscript{2} beyond this diradical character, BS wave function is the better choice. Fig.~\ref{fig:h2_simulation_results}b shows the time required to converge to the ground state for QITE under various bond lengths. Here, ``converge'' means that the QITE achieves chemical precision ($\Delta E = E_{\mathrm{QITE}} - E_{\mathrm{CAS\text{-}CI}} \le 1.0\ \mathrm{kcal\ mol^{-1}} (1/627.51 \ \text{Hartree})$).
QITE with RHF wave function takes a long time to converge as the bond length increases, while it is the opposite in the case of QITE with BS wave function. The background colors indicate recommended choice of starting wave function under varying bond lengths. 

The diradical character $y$ of the H$_2$ molecule during bond dissociation is shown in Fig.~\ref{fig:h2_simulation_results}c. BS-UHF converged to the RHF solution and $y$ is calculated to be 0 for bond length up to 1.2 Å.
Increasing the bond length further results in an increase of the diradical character, motivating the molecular geometry to have unpaired electrons in different orbitals.
At a bond length of around 3.0 Å, the diradical character is very close to 1. This is the region where the energy gap between the ground state and the first excited state is close to 0. 

\subsection{P4 cluster using QITE}
\begin{figure}
    \centering
        \includegraphics[width=1\columnwidth]{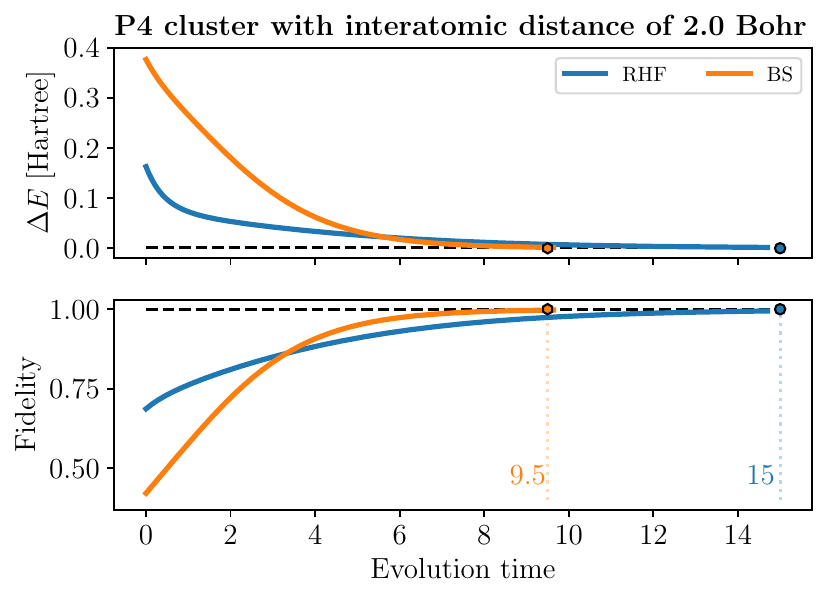}
    \caption{Comparison of the energy difference between the ground state energy calculated using CAS-CI and using HF and BS wave function inspired QITE with unitary approximation in the top panel. Bottom panel shows the fidelity of QITE evolved wave function with respect to the CAS-CI wave function for tetrahydrogen cluster with H--H interatomic distance of 2.0 Bohr.}
    \label{fig:figure5}
\end{figure}

Tetrahydrogen cluster shows a very strong open shell character at the square geometry with the bond length of 2.0 Bohr. In this geometry, even considering the domain size of 6 qubits, the HF wave function assisted QITE takes longer time to converge to the ground state. Fig.~\ref{fig:figure5} confirms the faster convergence of BS wave function assisted QITE with the same domain size during numerical simulation. It compares the fidelities of the wave funciton with BS and HF assisted QITE with CAS-CI wave function during the time evolution. Though HF assisted QITE has higher fidelity as the evolution initiates, BS assisted QITE rises above it and achieves chemical precision early. The chemical precision was attained in \(\approx 950\) iteration steps with BS assisted QITE, while it took \(\approx 1500\)  steps with HF assisted QITE.
The conditions taken for the simulation are described in Section~\ref{sec:computational conditions}.

\section{Scaling for large molecular systems}
For smaller molecules like H$_2$, it is fine to run these approximation methods without the need of partial Hamiltonians considering locality. For a nearest neighbor local Hamiltonian on a cubic lattice with dimension $d$, the domain size $D$ is bounded by $\mathcal{O}(C^d)$, where $C$ is the maximum number of qubits with finite correlation. Thus, considering a molecular Hamiltonian without locality, we take domain size and maximum number of qubits with finite correlation as the total number of qubits in the Hamiltonian for exact imaginary time evolution. The number of measurements and storage needed at a time stamp is bounded by $e^{{\mathcal{O}(C^d)}}$ since each unitary acts on $\mathcal{O}(C^d)$ qubits. Hence, the space and time complexity increases with $e^{\mathcal{O}(C^d)}$. However, if we localize all molecular orbitals within the active space using conventional orbital localization methods~\cite{r38Boys1966},~\cite{r39Pipek1989} and adopt Hamiltonian term truncation based on qubit operator locality~\cite{r40sugisaki2024}, a smaller domain size $D_{\mathrm{BS}}$ for BS inspired QITE is expected to perform better than HF inspired QITE with domain size $D_{\mathrm{HF}}$ such that $D_{\mathrm{BS}} \leq D_{\mathrm{HF}} < C$. This has not been examined in the current scope and is set for future works. Besides, the cost of preparation of a BS state is the same as that of a HF state in a quantum computer. Although the implementation in this paper is based on statevector simulation, implementing it in a real quantum device causes more depth of the circuit as the number of iterations increases. Thus, we might find a significant difference in the depth of the circuit while performing gate-level implementation. Since the BS inspired QITE is proven to require less number of iterations than the RHF inspired QITE for a particular region of strong open shell character, we can significantly reduce the circuit depth of the conventional QITE circuit.  

Together with these advantages comes some overhead in the implementation of the penalty operator. In addition to the computational cost of the penalty operator, it introduces new terms in the Hamiltonian, increasing the depth of the quantum circuit for each iteration. However, this overhead is taken over by a lesser number of iterations for BS inspired QITE. 
\begin{figure}[t]
    \centering
    \includegraphics[width=0.95\columnwidth]{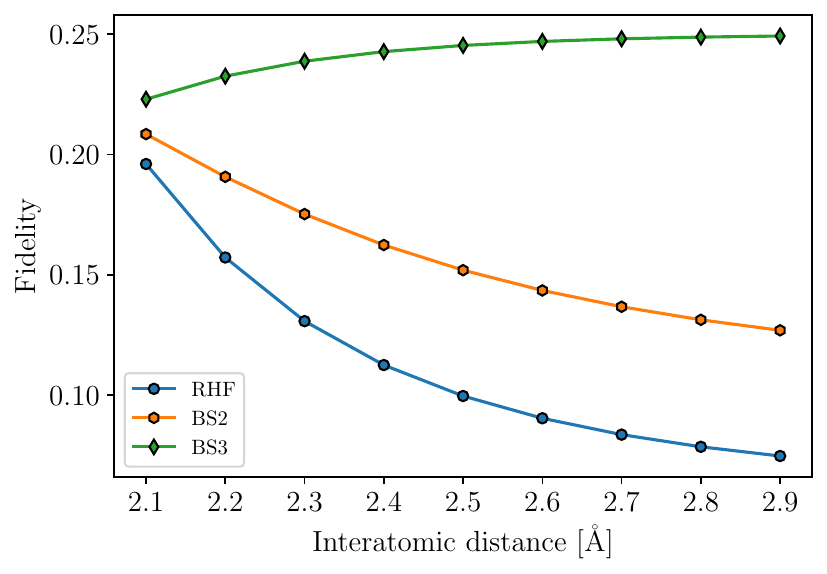}
    \caption{Initial fidelities of the wave functions against the CAS-CI wave function.}
    \label{fig:n2_initial_fidelities}
\end{figure}

In addition to this, the initial fidelity of these BS wave functions against the CAS-CI wave function at the corresponding bond length is greater than RHF wave function during the bond dissociation process as shown in Fig.~\ref{fig:n2_initial_fidelities}.
We considered the case of a N$_2$ molecule with six spins to study this behavior. As the molecule approaches bond dissociation limit, the initial fidelity of BS3 wave function approaches 0.25, while that of RHF decreases to 0.05. Thus, the BS wave function is itself a better approximation for open shell system than a RHF wave function. However, for the case of H$_2$ molecule with two spins and P4 cluster with four spins, HF wave function was found to have larger initial overlap than BS wave function even near the bond dissociation limit. We can understand that for a larger system executing open shell character, the BS wave function as initial wave function can become superior in cases where the initial overlap of the wave function with the CAS-CI values is crucial.

\section{Conclusion}
In this work, we studied the diradical character of the H$_2$ and N$_2$ molecules and a P4 cluster and proposed a BS wave function as a starting wave function in QITE in conjunction with the $\hat{\mathbf{S}}^2$ penalty operator to exploit the advantage of using a BS wave function. Performance of the proposed approach was studied with direct matrix exponentiation for ITE on N$_2$ molecule bond dissociation. Then the bond dissociation of the hydrogen molecule was studied under BS wave function with a unitary approximated QITE. The BS wave function inspired QITE performs better than a HF wave function inspired QITE as the system approaches the bond dissociation limit. We can conclude that, for the H$_2$ molecule, BS wave function is a better choice when the molecule starts to show open shell character after exhibiting the diradical character 0.21 and corresponding bond length of 1.56 Å. Prior to that, near the region of equilibrium, we recommend using a HF wave function. 

H$_2$ molecule was taken for this study for its open shell character at bond dissociation and the smaller size of the molecule. We can also use the BS inspired approach for larger system considering a proper method of unitary approximation.

\appendix
\section{Derivations}

\subsection{Branching diagram functions}
\label{sec:appA}

\begin{figure}[t]
    \centering
        \includegraphics[width=1\columnwidth]{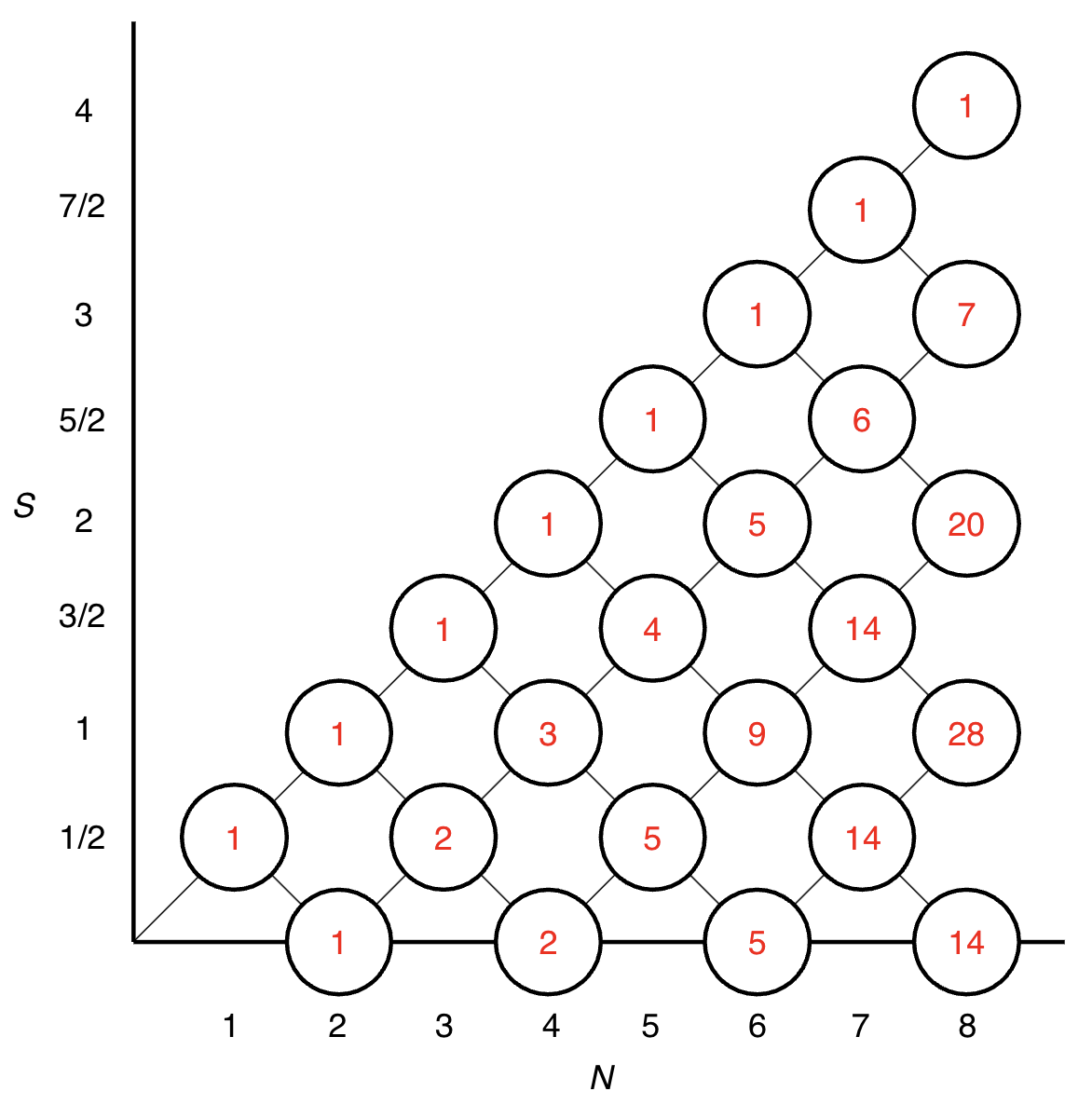}
    \caption{Branching diagram for the construction of spin eigenfunctions.}
    \label{fig:subfig6}
\end{figure}

The branching diagram shown in Fig.~\ref{fig:subfig6} is a blueprint for constructing spin eigenfunctions~\cite{r5Pauncz2000}. It consists of circles on the vertices and diagonal upward ($45^\circ$) and downward ($-45^\circ$) lines connecting vertices. The upward (downward) line represents adding an electron spin to increase (decrease) spin quantum number. The number in the circle represents the number of possible pathways to arrive the vertex and it corresponds to the number of spin eigenfunctions for given $N$ and $S$. In this work, we use the branching diagram to calculate the expansion coefficients of the spin eigenfunctions for the BS wave function. In fact, the BS wave function of an $N$-spin system with magnetic quantum number $M$ can be written as the linear combinations of the spin eigenfunctions as follows: 
\begin{eqnarray}
    |\mathrm{BS}(N, M)\rangle = \sum_j \sqrt\frac{g_{N,j}}{G_N} |N,S=j, M\rangle .
\end{eqnarray}

Here, $g_{N,j}$ is the number of spin eigenfunctions (values in the circle in the branching diagram) of $N$-spin system with $S=j$, and $G_N = \sum_k g_{N,k}$. Using this equation, we can easily derive the expanded form of the BS wave function in terms of spin eigenfunctions. For example, for two-spin system, the BS wave function $|\alpha\beta\rangle$ can be written as the linear combination of $|2,1,0\rangle $ and $|2,0,0\rangle $ states, and the coefficients are $\frac{1}{\sqrt{2}}$ for both. Therefore, 
\begin{align}
    &\quad |\alpha\beta\rangle = \frac{1}{\sqrt{2}}|2,0,0\rangle + \frac{1}{\sqrt{2}}|2,1,0\rangle .
\end{align}
A six-spin BS wave function of $M = 0$ can be derived as: 
\begin{align}
    |\alpha\alpha\alpha\beta\beta\beta\rangle &= \frac{1}{\sqrt{20}}|6,3,0\rangle +\frac{1}{{2}}|6,2,0\rangle + \notag \\
    &\quad \frac{3}{\sqrt{20}}|6,1,0\rangle + \frac{1}{{2}}|6,0,0\rangle.    
\end{align}

\bibliographystyle{IEEEtran.bst}
\bibliography{IEEEabrv, references}

\end{document}

%% file: custom-commands.tex
\usepackage{ifthen}
\newboolean{ShowComments}
\setboolean{ShowComments}{true}  
\ifthenelse{\boolean{ShowComments}}%
	{
		\newcommand{\ColorComment}[3]{%
				{\colorbox{#1}{\color{White}   \textsf{\textbf{#2}}} \textcolor{#1}{#3}}}

	}%
	{
		\newcommand{\ColorComment}[3]{}

	}%

\definecolor{andreascolor}{rgb}{0,0.5,0}
\definecolor{stephancolor}{RGB}{255,127,80}
\definecolor{naphancolor}{RGB}{165,6,245}
\definecolor{lpcolor}{RGB}{0, 0, 255}
\definecolor{nathancolor}{RGB}{48, 194, 174}
\definecolor{johncolor}{RGB}{178, 152, 36}
\definecolor{reubencolor}{RGB}{178, 152, 36}
\definecolor{pawancolor}{RGB}{0, 128, 128}

\newcommand*{\SavedEqref}{}
\let\SavedEqref\eqref
\renewcommand*{\eqref}[1]{%
  \begingroup
    \hypersetup{
      linkcolor=CornflowerBlue,
      linkbordercolor=CornflowerBlue,
    }%
    \SavedEqref{#1}%
  \endgroup
}

%% file: main.bbl
\begin{thebibliography}{10}
\providecommand{\url}[1]{#1}
\csname url@samestyle\endcsname
\providecommand{\newblock}{\relax}
\providecommand{\bibinfo}[2]{#2}
\providecommand{\BIBentrySTDinterwordspacing}{\spaceskip=0pt\relax}
\providecommand{\BIBentryALTinterwordstretchfactor}{4}
\providecommand{\BIBentryALTinterwordspacing}{\spaceskip=\fontdimen2\font plus
\BIBentryALTinterwordstretchfactor\fontdimen3\font minus \fontdimen4\font\relax}
\providecommand{\BIBforeignlanguage}[2]{{%
\expandafter\ifx\csname l@#1\endcsname\relax
\typeout{** WARNING: IEEEtran.bst: No hyphenation pattern has been}%
\typeout{** loaded for the language `#1'. Using the pattern for}%
\typeout{** the default language instead.}%
\else
\language=\csname l@#1\endcsname
\fi
#2}}
\providecommand{\BIBdecl}{\relax}
\BIBdecl

\bibitem{r24Aspuru-Guzik2005}
A.~Aspuru-Guzik, A.~D. Dutoi, P.~J. Love, and M.~Head-Gordon, ``Simulated quantum computation of molecular energies,'' \emph{Science}, vol. 309, no. 5741, pp. 1704--1707, 2005, \href{https://dx.doi.org/10.1126/science.1113479}{doi:10.1126/science.1113479}.

\bibitem{r12Sala2025}
L.~C. Sala, P.~A.-A. L. D.~I. S.L., and M.~M. Lab, ``Efficient protein ground state energy computation via fragmentation and reassembly,'' 2025, \href{https://dx.doi.org/10.48550/arXiv.2501.03766}{doi:10.48550/arXiv.2501.03766}.

\bibitem{r13Kim2022}
I.~H. Kim, Y.-H. Liu, S.~Pallister, W.~Pol, S.~Roberts, and E.~Lee, ``Fault-tolerant resource estimate for quantum chemical simulations: Case study on {L}i-ion battery electrolyte molecules,'' \emph{Phys. Rev. Res.}, vol.~4, p. 023019, 2022, \href{https://dx.doi.org/10.1103/PhysRevResearch.4.023019}{doi:10.1103/PhysRevResearch.4.023019}.

\bibitem{r22Kitaev1995}
A.~Y. Kitaev, ``Quantum measurements and the abelian stabilizer problem,'' 1995, \href{https://dx.doi.org/10.48550/arXiv.quant-ph/9511026}{doi:10.48550/arXiv.quant-ph/9511026}.

\bibitem{r23Abrams1999}
D.~S. Abrams and S.~Lloyd, ``Quantum algorithm providing exponential speed increase for finding eigenvalues and eigenvectors,'' \emph{Phys. Rev. Lett.}, vol.~83, pp. 5162--5165, 1999, \href{https://dx.doi.org/10.1103/PhysRevLett.83.5162}{doi:10.1103/PhysRevLett.83.5162}.

\bibitem{r14Belaloui2024}
N.~E. Belaloui \emph{et~al.}, ``Ground state energy estimation on current quantum hardware through the variational quantum eigensolver: A comprehensive study,'' 2024, \href{https://dx.doi.org/10.48550/arXiv.2412.02606}{doi:10.48550/arXiv.2412.02606}.

\bibitem{r44Huggins2022}
W.~J. Huggins, B.~A. O’Gorman, N.~C. Rubin, D.~R. Reichman, R.~Babbush, and J.~Lee, ``Unbiasing fermionic quantum {M}onte {C}arlo with a quantum computer,'' \emph{Nature}, vol. 603, no. 7901, pp. 416--420, 2022, \href{https://dx.doi.org/10.1038/s41586-021-04351-z}{doi:10.1038/s41586-021-04351-z}.

\bibitem{r4Motta2020}
M.~Motta \emph{et~al.}, ``Determining eigenstates and thermal states on a quantum computer using quantum imaginary time evolution,'' \emph{Nature Physics}, vol.~16, pp. 205--210, 2020, \href{https://dx.doi.org/10.1038/s41567-020-0798-8}{doi:10.1038/s41567-020-0798-8}.

\bibitem{r1Sugisaki2016}
K.~Sugisaki \emph{et~al.}, ``Quantum chemistry on quantum computers: A polynomial-time quantum algorithm for constructing the wave functions of open-shell molecules,'' \emph{The Journal of Physical Chemistry A}, vol. 120, pp. 6459--6466, 2016, \href{https://dx.doi.org/10.1021/acs.jpca.6b04932}{doi:10.1021/acs.jpca.6b04932}.

\bibitem{r25Sugisaki2019}
K.~Sugisaki, S.~Nakazawa, K.~Toyota, K.~Sato, D.~Shiomi, and T.~Takui, ``Quantum chemistry on quantum computers: A method for preparation of multiconfigurational wave functions on quantum computers without performing post-{H}artree-{F}ock calculations,'' \emph{ACS Cent. Sci.}, vol.~5, pp. 167--175, 2019, \href{https://dx.doi.org/10.1021/acscentsci.8b00788}{doi:10.1021/acscentsci.8b00788}.

\bibitem{r26lee2023}
S.~Lee \emph{et~al.}, ``Evaluating the evidence for exponential quantum advantage in ground-state quantum chemistry,'' \emph{Nature Communications}, vol.~14, no.~1, p. 1952, 2023, \href{https://dx.doi.org/10.1038/s41467-023-37587-6}{doi:10.1038/s41467-023-37587-6}.

\bibitem{r3Sugisaki2022}
K.~Sugisaki, K.~Toyota, K.~Sato, D.~Shiomi, and T.~Takui, ``Adiabatic state preparation of correlated wave functions with nonlinear scheduling functions and broken-symmetry wave functions,'' \emph{Communications Chemistry}, vol.~5, no.~1, p.~84, 2022, \href{https://dx.doi.org/10.1038/s42004-022-00701-8}{doi:10.1038/s42004-022-00701-8}.

\bibitem{r27Coronado2020}
E.~Coronado, ``Molecular magnetism: from chemical design to spin control in molecules, materials and devices,'' \emph{Nature Reviews Materials}, vol.~5, pp. 87--104, 2020, \href{https://dx.doi.org/10.1038/s41578-019-0146-8}{doi:10.1038/s41578-019-0146-8}.

\bibitem{r28nakano2015}
M.~Nakano and B.~Champagne, ``Theoretical design of open-shell singlet molecular systems for nonlinear optics,'' \emph{The Journal of Physical Chemistry Letters}, vol.~6, no.~16, pp. 3236--3256, 2015, \href{https://dx.doi.org/10.1021/acs.jpclett.5b00956}{doi:10.1021/acs.jpclett.5b00956}.

\bibitem{r29umena2011}
Y.~Umena, K.~Kawakami, J.-R. Shen, and N.~Kamiya, ``Crystal structure of oxygen-evolving photosystem {II} at a resolution of 1.9 Å,'' \emph{Nature}, vol. 473, no. 7345, pp. 55--60, 2011, \href{https://dx.doi.org/10.1038/nature09913}{doi:10.1038/nature09913}.

\bibitem{r30spatzal2016}
T.~Spatzal \emph{et~al.}, ``Nitrogenase {FeMoco} investigated by spatially resolved anomalous dispersion refinement,'' \emph{Nature Communications}, vol.~7, no.~1, p. 10902, 2016, \href{https://dx.doi.org/10.1038/ncomms10902}{doi:10.1038/ncomms10902}.

\bibitem{r10Sugisaki2022}
K.~Sugisaki, T.~Kato, Y.~Minato, K.~Okuwaki, and Y.~Mochizuki, ``Variational quantum eigensolver simulations with the multireference unitary coupled cluster ansatz: a case study of the {C}$_\mathrm{2v}$ quasi-reaction pathway of beryllium insertion into a {H}$_2$ molecule,'' \emph{Phys. Chem. Chem. Phys.}, vol.~24, pp. 8439--8452, 2022, \href{https://dx.doi.org/10.1039/D1CP04318H}{doi:10.1039/D1CP04318H}.

\bibitem{preskill2018nisq}
J.~Preskill, ``Quantum {C}omputing in the {NISQ} era and beyond,'' \emph{{Quantum}}, vol.~2, p.~79, Aug. 2018, \href{https://dx.doi.org/10.22331/q-2018-08-06-79}{doi:10.22331/q-2018-08-06-79}.

\bibitem{r31anand2022}
A.~Anand \emph{et~al.}, ``A quantum computing view on unitary coupled cluster theory,'' \emph{Chemical Society Reviews}, vol.~51, no.~5, pp. 1659--1684, 2022, \href{https://dx.doi.org/10.1039/D1CS00932J}{doi:10.1039/D1CS00932J}.

\bibitem{r32kremenetski2021}
V.~Kremenetski, C.~Mejuto-Zaera, S.~J. Cotton, and N.~M. Tubman, ``Simulation of adiabatic quantum computing for molecular ground states,'' \emph{The Journal of Chemical Physics}, vol. 155, no.~23, p. 234106, 12 2021, \href{https://dx.doi.org/10.1063/5.0060124}{doi:10.1063/5.0060124}.

\bibitem{r37Paldus1993}
J.~Paldus, P.~Piecuch, L.~Pylypow, and B.~Jeziorski, ``Application of {H}ilbert-space coupled-cluster theory to simple ({${\mathrm{H}}_{2}$${)}_{2}$} model systems: Planar models,'' \emph{Phys. Rev. A}, vol.~47, no.~4, pp. 2738--2782, April 1993, \href{https://dx.doi.org/10.1103/PhysRevA.47.2738}{doi:10.1103/PhysRevA.47.2738}.

\bibitem{r43Wick1954}
G.~C. Wick, ``Properties of {B}ethe-{S}alpeter wave functions,'' \emph{Physical Review}, vol.~96, no.~4, pp. 1124--1134, Nov 1954, \href{https://dx.doi.org/10.1103/PhysRev.96.1124}{doi:10.1103/PhysRev.96.1124}.

\bibitem{r17Liu2020ProbabilisticNG}
T.~Liu, J.-G. Liu, and H.~Fan, ``Probabilistic nonunitary gate in imaginary time evolution,'' \emph{Quantum Information Processing}, vol.~20, 2020, \href{https://dx.doi.org/10.1007/s11128-021-03145-6}{doi:10.1007/s11128-021-03145-6}.

\bibitem{r18Ammar2017}
A.~Daskin and S.~Kais, ``An ancilla-based quantum simulation framework for non-unitary matrices,'' vol.~16, no.~1, p. 1–17, 2017, \href{https://dx.doi.org/10.1007/s11128-016-1452-3}{doi:10.1007/s11128-016-1452-3}.

\bibitem{r19Gily2019}
A.~Gily\'{e}n, Y.~Su, G.~H. Low, and N.~Wiebe, ``Quantum singular value transformation and beyond: exponential improvements for quantum matrix arithmetics,'' in \emph{Proceedings of the 51st Annual ACM SIGACT Symposium on Theory of Computing}, ser. STOC 2019.\hskip 1em plus 0.5em minus 0.4em\relax New York, NY, USA: Association for Computing Machinery, 2019, p. 193–204, \href{https://dx.doi.org/10.1145/3313276.3316366}{doi:10.1145/3313276.3316366}.

\bibitem{r33YAMAGUCHI1975}
K.~Yamaguchi, ``The electronic structures of biradicals in the unrestricted {H}artree-{F}ock approximation,'' \emph{Chemical Physics Letters}, vol.~33, no.~2, pp. 330--335, 1975, \href{https://dx.doi.org/10.1016/0009-2614(75)80169-2}{doi:10.1016/0009-2614(75)80169-2}.

\bibitem{r34shoji2006}
M.~Shoji \emph{et~al.}, ``A general algorithm for calculation of {H}eisenberg exchange integrals {J} in multispin systems,'' \emph{Chemical Physics Letters}, vol. 432, no.~1, pp. 343--347, 2006, \href{https://dx.doi.org/10.1016/j.cplett.2006.10.023}{doi:10.1016/j.cplett.2006.10.023}.

\bibitem{r35Zahariev2023}
F.~Zahariev \emph{et~al.}, ``\BIBforeignlanguage{English}{The general atomic and molecular electronic structure system ({GAMESS}): Novel methods on novel architectures},'' \emph{\BIBforeignlanguage{English}{Journal of Chemical Theory and Computation}}, vol.~19, no.~20, pp. 7031--7055, October 2023, \href{https://dx.doi.org/10.1021/acs.jctc.3c00379}{doi:10.1021/acs.jctc.3c00379}.

\bibitem{r36McClean_2020}
J.~R. McClean \emph{et~al.}, ``Open{F}ermion: The electronic structure package for quantum computers,'' \emph{Quantum Science and Technology}, vol.~5, no.~3, p. 034014, June 2020, \href{https://dx.doi.org/10.1088/2058-9565/ab8ebc}{doi:10.1088/2058-9565/ab8ebc}.

\bibitem{Doehnert1980}
D.~Doehnert and J.~Koutecky, ``Occupation numbers of natural orbitals as a criterion for biradical character. different kinds of biradicals,'' \emph{Journal of the American Chemical Society}, vol. 102, no.~6, pp. 1789--1796, 1980, \href{https://dx.doi.org/10.1021/ja00526a005}{doi:10.1021/ja00526a005}.

\bibitem{r38Boys1966}
S.~F. Boys, ``Quantum science of atoms, molecules, and solids,'' in \emph{Quantum Science of Atoms, Molecules, and Solids}, P.~O. Lowdin, Ed.\hskip 1em plus 0.5em minus 0.4em\relax New York: Academic Press, 1966, pp. 253--262.

\bibitem{r39Pipek1989}
J.~Pipek and P.~G. Mezey, ``A fast intrinsic localization procedure applicable for ab initio and semiempirical linear combination of atomic orbital wave functions,'' \emph{The Journal of Chemical Physics}, vol.~90, no.~9, pp. 4916--4926, May 1989, \href{https://dx.doi.org/10.1063/1.456588}{doi:10.1063/1.456588}.

\bibitem{r40sugisaki2024}
K.~Sugisaki, S.~Kanno, T.~Itoko, R.~Sakuma, and N.~Yamamoto, ``Hamiltonian simulation-based quantum-selected configuration interaction for large-scale electronic structure calculations with a quantum computer,'' 2024, \href{https://dx.doi.org/10.48550/arXiv.2412.07218}{doi:10.48550/arXiv.2412.07218}.

\bibitem{r5Pauncz2000}
R.~Pauncz, \emph{The Construction of Spin Eigenfunctions}.\hskip 1em plus 0.5em minus 0.4em\relax Kluwer Academic/Plenum Publishers, 2000.

\end{thebibliography}
